\documentclass[conference]{IEEEtran}
\IEEEoverridecommandlockouts

\usepackage{fancyhdr}
\usepackage{amsmath,amssymb,amsfonts}
\usepackage{algorithmic}
\usepackage{graphicx}
\usepackage{textcomp}
\usepackage{xcolor}
\usepackage{svg}
\usepackage{multirow}
\usepackage{indentfirst}
\usepackage{array}
\usepackage{tabularray}
\usepackage{pgfplots}
\usepackage[linesnumbered]{algorithm2e}
\usepackage{calc}
\usepackage{subcaption}
\usepackage{enumitem}
\usepackage{setspace}
\usepackage{pgfplots}
\usepackage{makecell}
\usepackage{booktabs} 
\usepackage{array}
\RestyleAlgo{ruled}
\usepackage{amssymb} 
\usepackage{pifont}  
\usepackage{fancyhdr} 
\usepackage{lipsum}   
\usepackage{fancyhdr}
\usepackage{amsmath,amssymb,amsfonts}
\usepackage{algorithmic}
\usepackage{graphicx}
\usepackage{textcomp}
\usepackage{xcolor}
\usepackage{svg}
\usepackage{multirow}
\usepackage{indentfirst}
\usepackage{array}
\usepackage{tabularray}
\usepackage{pgfplots}
\usepackage[linesnumbered]{algorithm2e}
\usepackage{calc}
\usepackage{subcaption}
\usepackage{enumitem}
\usepackage{setspace}
\usepackage{pgfplots}
\usepackage{fancyhdr} 
\usepackage{lipsum}   
\usepackage{marvosym}
\RestyleAlgo{ruled}

\SetKwInOut{KwIn}{Input}
\SetKwInOut{KwOut}{Output}
\def\BibTeX{{\rm B\kern-.05em{\sc i\kern-.025em b}\kern-.08em
    T\kern-.1667em\lower.7ex\hbox{E}\kern-.125emX}}
\begin{document}

\title{Extend IVerilog to Support Batch RTL Fault Simulation
\\
}
\author{\IEEEauthorblockN{
Jiaping Tang\textsuperscript{*}\textsuperscript{\dag}\textsuperscript{\ddag},
Jianan Mu\textsuperscript{*}\textsuperscript{\ddag}\textsuperscript{\Letter}, 
Zizhen Liu\textsuperscript{*}\textsuperscript{\ddag}, 
Zhiteng Chao\textsuperscript{*}\textsuperscript{\ddag}, 
Jing Ye\textsuperscript{*}\textsuperscript{\dag}\textsuperscript{\ddag},
Huawei Li\textsuperscript{*}\textsuperscript{\dag}\textsuperscript{\ddag}\textsuperscript{\Letter} }
\IEEEauthorblockA{
\textsuperscript{*}\textit{State Key Lab of Processors, Institute of Computing Technology, Chinese Academy of Sciences, Beijing, China} \\
\textsuperscript{\dag}\textit{University of Chinese Academy of Sciences, Beijing, China}\\
\textsuperscript{\ddag}\textit{CASTEST Co., Ltd., Beijing, China}\\
\{tangjiaping22s, mujianan, liuzizhen, chaozhiteng21, yejing, lihuawei\}@ict.ac.cn}
}
\maketitle

\begin{abstract}
The advancement of functional safety has made RTL-level fault simulation increasingly important to achieve iterative efficiency in the early stages of design and to ensure compliance with functional safety standards. In this paper, we extend IVerilog to support batch RTL fault simulation and integrate the event-driven algorithm and the concurrent fault simulation algorithm. Comparative experiments with a state-of-the-art commercial simulator and an open-source RTL fault simulator demonstrate that our simulator achieves a performance improvement of 2.2$\times$ and 3.4$\times$, respectively.
\end{abstract}

\section{Introduction}

As automobiles evolve, ensuring the functional safety of critical components becomes essential.
Functional safety verification demands extensive RTL fault simulations, which are highly time-consuming and have become a bottleneck in practical development \cite{long_em, long_2}.

Recently, some works are proposed to extend the functionality of open-soruce simulators, such as verilator \cite{verilator}, to implement fault simulation \cite{FSim_1, FSim_2}. 
Kaja et al. \cite{FSim_1} extended Verilator by encapsulating each signal to an object and overwriting the assign operator of the object to achieve fault injection capabilities. 
Geier et al. \cite{FSim_2} proposed an open-source tool vRTLmod, which takes the Verilator output and automatically adds the fault injection capability at flip-flops.
However, classic optimization methods, such as concurrent or event-driven approaches, have not been sufficiently integrated into these fault simulators. Due to Verilator's inherent cycle-based and compiled characteristics, it is even impractically impossible to integrate them. Therefore, the injection and simulation of a single fault at a time are time-consuming. 

 
The IVerilog \cite{iverilog} event-driven simulation algorithm makes it highly suitable for fault simulation. Due to the limited fault propagation capability, event-driven effectively avoids simulating across the entire circuit. To take advantage of this, we extended IVerilog to enhance its fault simulation capabilities.  
To maximize reuse of logic simulation results, we integrate a concurrent fault simulation algorithm into IVerilog. This integration enables simultaneous and independent batch simulation of multiple faults.  

\section{RTL fault simulation}\label{section:framework}
RTL code comprises two components: the RTL netlist and behavioral descriptions or behavioral codes. The RTL netlist represents connections between primitives and variables, while the behavioral descriptions define \textit{initial} and \textit{always} blocks in RTL. Similarly, IVerilog provides corresponding data structures for these components. In the following sections, we will explain the implementation of fault simulation for both the RTL netlist and behavioral descriptions.

\subsection{Fault Simulation on the RTL Netlist}\label{netlist}

The concurrent fault simulation in the RTL netlist is implemented encompassing five main steps: allocation of bad gates for fault effect storage, fault injection, evaluation, identification of visibility for bad gates, and propagation of good and bad events. 
The allocation of bad gates involves allocating memory for each node in the RTL netlist based on the injected faults. 
Fault injection attaches faults to the corresponding bad gates, leaving them null if no faults are present. 
Evaluation determines a node's output based on its inputs and type. When faults are present, the output is filtered through the faults to generate the final result.
The visibility of bad gates is identified by comparing the outputs of good and bad gates; if they differ, the bad gate is deemed visible, and a corresponding bad event is generated. 
Finally, propagation of good and bad events involves sending new good or bad values to successor nodes and activating their evaluation. 
Some events may activate the execution of behavioral codes. More details will be elaborated in Section \ref{behavioral}.

\begin{table*}[h]
\centering
\setlength{\abovecaptionskip}{0cm}
\caption{Performance comparison with various RTL fault simulators}

\label{test:fsim_time}
\resizebox{0.94\linewidth}{!}{
\begin{tabular}{c|c|c|c|c|c|c|c}
\hline
\multirow{2}{*}{\textbf{Design}} &
\multirow{2}{*}{\textbf{Stimulus cycles}} &
\multirow{2}{*}{\textbf{\#Cells}} &
\multirow{2}{*}{\textbf{\#Wire faults}} &
\multirow{2}{*}{\textbf{Ours(s)}} &
\multicolumn{3}{c}{\textbf{Performance: time(s) / Speedup(others$/$ours)}}
 \\
\cline{6-8}
& & & & & \textbf{Verilator\_Fsim} & \textbf{Iverilog\_Fsim} & \textbf{Z01X} \\

\hline
Uart & 169k & 283 & 88 & \textbf{2.42} & 9.9 / (4.1$\times$) & 10 / (4.2$\times$) & 3 / (1.2$\times$)  \\
\hline
SDRAM Controller & 43k & 337 & 278 & 10 & 17.1 / (1.7$\times$) & 24.7 / (2.5$\times$) & \textbf{5 / (0.5$\times$)}   \\
\hline
Fixed Point Adder & 5.0k & 1591 & 384 & 0.55 & \textbf{0.51 / (0.9$\times$)} & 3.1 / (5.6$\times$) & 3 / (5.5$\times$)  \\
\hline
Fixed Point Multiplier & 5.0k & 17039 & 384 & \textbf{0.57} & 1.02  / (1.8$\times$) & 5.4 / (9.5$\times$) & 3 / (5.3$\times$)   \\
\hline
Float Point Adder & 60k & 2000 & 208 & \textbf{1.55} & 12.7  / (8.2$\times$)& 45.8 / (29.6$\times$) & 2 / (1.3$\times$)  \\
\hline
Float Point Multiplier &  60k & 4899 &  208 & 2.52  & 14.2 / (5.6$\times$) & 52.0 / (20.7$\times$) & \textbf{2 / (0.8$\times$)}  \\
\hline
Finite Impulse Responder Filter & 10k  & 566 & 626 & \textbf{2.84}   & 4.3 / (1.5$\times$) & 28.3  / (10.0$\times$)& 3 / (1.1$\times$) \\
\hline
Average speedup & \multicolumn{3}{c|}{-} & - & \textbf{3.4}$\times$& \textbf{20.5}$\times$& \textbf{2.23}$\times$ \\
\hline
\end{tabular}
}
 
\end{table*}

\subsection{Fault Simulation on Behavioral Descriptions}\label{behavioral}
In fault simulation, fault effects may propagate through behavioral nodes. The behavioral code reads fault effects from the RTL netlist, processes them, and writes the processed fault effects back to the RTL netlist. Since we support simultaneous and independent fault simulations, preventing interference between fault effects is crucial.

To address this, we create a separate instance for each fault in each behavioral description, ensuring that each instance handles only the corresponding fault effects. Specifically, a \textit{fault id} attribute is introduced to indicate how data should be read and written to the RTL netlist.
When triggered by posedge events, negedge events, or delay events, the behavioral code reads fault effects from the specified bad gate of the RTL netlist node based on the \textit{fault id}, processes the effects, and writes the results back to the bad gate indicated by \textit{fault id} of the corresponding RTL netlist node.

\subsection{A Case Study}

\begin{figure}[htbp]
\centerline{
\includegraphics[width=0.9\linewidth]{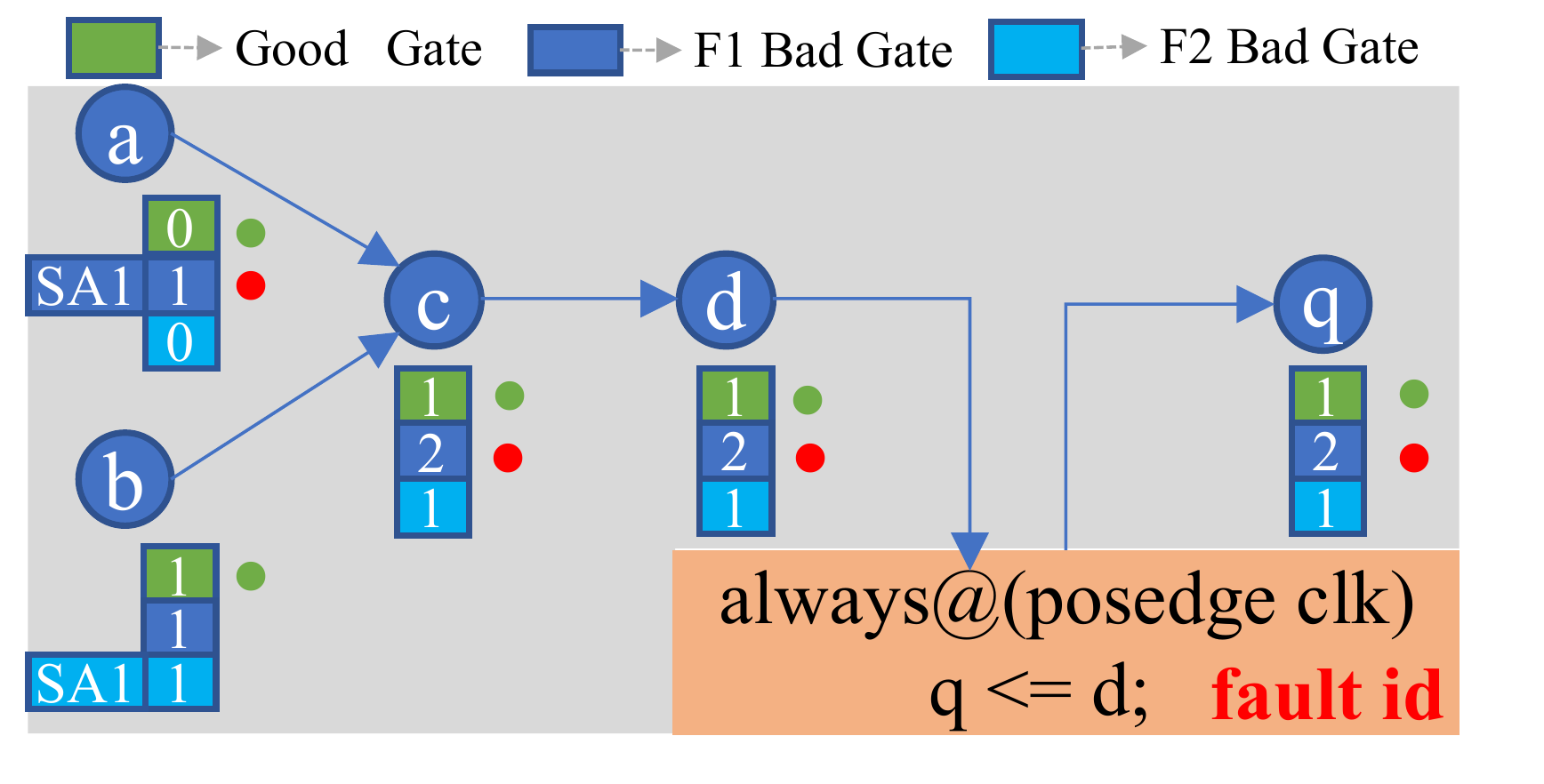}
}
\caption{Concurrent fault simulation on the RTL.
}
\label{netlist_fsim}
\end{figure}

 Figure \ref{netlist_fsim} shows an example of fault simulation on the RTL. 
The gray areas represent the RTL netlist, while the orange areas indicate the behavioral code. Green dots on good gates signify the presence of good events, while red dots on bad gates indicate bad events. SA1 faults are injected at nodes \textit{a} and \textit{b}, denoted as \textit{F1} and \textit{F2}, respectively. These faults are associated with the corresponding bad gates of nodes \textit{a} and \textit{b}.  

Assuming all initial values in the RTL netlist are \textit{x}, stimuli of 0 and 1 are applied to \textit{a} and \textit{b}, generating good events on these nodes, marked by green dots. The visibility of bad gates at \textit{a} and \textit{b} is then assessed. For the bad gate at \textit{a}, since the value of the bad gate associated with \textit{F1} is 1, which differs from the good gate value, a bad event is generated for \textit{F1}, marked by red dots. The same analysis applies to other nodes.

If a posedge event of clock occurs, it triggers the execution of behavioral code. First, the fault-free behavioral code executes, as its \textit{fault id} is set to \textit{invalid}. It reads the value from the good gate of node \textit{d} and writes this value back to the good gate of node \textit{q}.  
Next, the behavioral code for fault \textit{F1} executes. With a \textit{fault id = 0}, it reads the value from the first bad gate (index 0) of node \textit{d} and writes the fault effect back to the first bad gate of node \textit{q}. Since a new event is created at node \textit{q}, the propagation continues by processing node q and generating further events until no new events occur.

\section{Evaluation}\label{section:eveulation}


We compare three RTL fault simulators: the commercial tool Z01X, Iverilog\_Fsim with \textit{force} command support, and the open-source Verilator\_Fsim \cite{FSim_1}. Benchmarks from OpenCores \cite{OpenCores} cover IPs for memory, communication, processors, and DSPs. Stuck-at faults are generated for all wires.


Table \ref{test:fsim_time} presents a comparison of fault simulation performance between Verilator\_Fsim, Iverilog\_Fsim, Z01X, and our simulator across all designs. 
Compared with  Verilator\_Fsim, Iverilog\_Fsim, and Z01X, our simulator achieves a performance improvement of 3.4$\times$, 20.5$\times$, and 2.2$\times$, respectively.

\textbf{Compared with the commercial simulator.} Our simulator outperforms the commercial one on most designs, especially on the fixed point adder design for a speedup of 5.5$\times$ and on the fixed point multiplier design for a speedup of 5.3$\times$. 
However, in the case of the SDRAM controller design, our performance is inferior to that of Z01X. We observe that Z01X performs testability analysis before fault simulation, which probably assists in identifying faults and improving simulation efficiency. This may be a reason why our performance is inferior for the SDRAM controller design. 

\textbf{Compared with the open-source simulator.} Our runtime is consistently lower than that of the state-of-the-art open-source simulator, which is particularly prominent for the float point adder design. 
We can observe that although the performance of Verilator\_Fsim is significantly better than Iverilog\_Fsim in single-fault scenarios, by incorporating our proposed methods in this paper, such as batch fault simulation and event-driven simulation, our simulation performance can surpass that of Verilator\_Fsim.

\section{Conclusion}\label{section:conclusion}
In this paper, we extend IVerilog to support fault simulation with integrating concurrent fault simulation algorithm and event-driven algorithm. Compared to a state-of-the-art commercial simulator and an open-source RTL fault simulator, our simulator achieves a performance improvement of 2.2$\times$ and 3.4$\times$, respectively.

\vspace{12pt}

\end{document}